\documentclass[sigconf, review=false, anonymous=false]{acmart} 
\AtBeginDocument{%
  \providecommand\BibTeX{{%
    \normalfont B\kern-0.5em{\scshape i\kern-0.25em b}\kern-0.8em\TeX}}}

\setcopyright{acmcopyright}
\copyrightyear{2022}
\acmYear{2022}
\acmDOI{xx.xxxx/xxxxxxx.xxxxxxx}

\usepackage[utf8]{inputenc}
\let\proof\relax

\usepackage{amsmath}
\usepackage{enumerate}
\usepackage{amsfonts}
\usepackage[normalem]{ulem}
\usepackage{color}
\usepackage{xcolor}
\usepackage{comment}
\usepackage{subcaption}
\usepackage{mathtools}
\usepackage{url}
\usepackage[ruled,linesnumbered]{algorithm2e}
\usepackage{xfrac}
\usepackage{enumitem}

\newcommand\given[1][]{\:#1\vert\:}
\let\textapprox\relax
\newcommand{\textapprox}{\raisebox{0.5ex}{\texttildelow}}
\newcommand{\proof}[1]{ \noindent \textit{Proof}. #1\qed }

\DeclareMathOperator{\EX}{\mathbb{E}}%
\usepackage{amsthm}
\newtheorem{problem}[]{Problem}
\newtheorem{theorem}{Theorem}[section]

\newtheorem{define}[theorem]{Definition}

\begin{document}
\title{Elliptical Slice Sampling for Probabilistic Verification of Stochastic Systems with Signal Temporal Logic Specifications}

\author{Guy Scher}
 \email{gs679@cornell.edu}
 \affiliation{%
   \institution{Sibley School of Mechanical and Aerospace Engineering, Cornell University}
   \city{Ithaca}
   \state{NY}
   \country{USA}
 }

\author{Sadra Sadraddini}
 \email{sadra@dexai.com}
 \affiliation{%
   \institution{Dexai Robotics}
   \city{Boston}
   \state{MA}
   \country{USA}
 }
\author{Russ Tedrake}
 \email{russt@mit.edu}
 \affiliation{%
   \institution{Computer Science and Artificial Intelligence Laboratory  (CSAIL), Massachusetts Institute of Technology}
   \city{Cambridge}
   \state{MA}
   \country{USA}
 }

\author{Hadas Kress-Gazit}
 \email{hadaskg@cornell.edu}
 \affiliation{%
   \institution{Sibley School of Mechanical and Aerospace Engineering, Cornell University}
   \city{Ithaca}
   \state{NY}
   \country{USA}
 }

\begin{abstract}
Autonomous robots typically incorporate complex sensors in their decision-making and control loops. These sensors, such as cameras and Lidars, have imperfections in their sensing and are influenced by environmental conditions.
In this paper, we present a method for probabilistic verification of linearizable systems with Gaussian and Gaussian mixture noise models (e.g. from perception modules, machine learning components). We compute the probabilities of task satisfaction under Signal Temporal Logic (STL) specifications, using its robustness semantics, with a Markov Chain Monte-Carlo slice sampler. As opposed to other techniques, our method avoids over-approximations and double-counting of failure events. Central to our approach is a method for efficient and rejection-free sampling of signals from a Gaussian distribution such that satisfy or violate a given STL formula. We show illustrative examples from applications in robot motion planning. 
\end{abstract}



\maketitle

\section{Introduction}
\label{sec:introduction}

\noindent To deploy autonomous robots, such as self-driving cars or assistive robots, we seek formal guarantees that they can operate safely and reliably. %
Providing such guarantees is challenging due to the sheer amount of non-determinism in the world including noisy sensors, uncontrolled environment (humans, other robots) and different environment conditions (such as lighting, occlusions, etc.). Modern systems also include machine learning components \cite{dean2020robust} that can contribute to uncertainty since they might be deployed in different settings than the ones they were trained on.

Sensors, from proprioceptive ones that sense the robot's internal values such as speed or joint angles, to exteroceptive ones that sense the environment such as range finders and cameras are usually modeled with errors coming from a Gaussian distribution or bounded noises. The system designer needs to reason about the likelihood that the system will successfully perform the task and re-design it if needed. The general approach is to find all the states (e.g. the robot's positions) that the robot may reach under all circumstances, i.e. the ``reachable set'', and reason about the safety and task completion.
Testing with hardware is limiting, impractical and intractable because of the variability of tests and environmental conditions. Finding rigorous formal mathematical guarantees is usually infeasible for complicated systems performing complex tasks. Verifying systems using simulations may be the only way, but they also suffer from long computation times, especially when searching for rare and hard to find events~\cite{chou2018using,o2018scalable,yu2018safe}.

Several techniques exist for verifying systems with uncertainty in the literature. Imposing hard constraints on the state will always result in violation when dealing with unbounded non-determinism such as the Gaussian noise model. As such, it makes sense to describe the constraints with the probability of satisfying them - probabilistic state constraints. One common approach to verifying such robotic systems is with \textbf{chance constraints}~\cite{du2011probabilistic,blackmore2009convex}. In these formulations, it is common to do risk allocation and use Boole's inequality, which allocates the level of uncertainty for each constraint component~\cite{ono2008efficient}, or use ellipsoidal approximations. However, both are considered to be conservative~\cite{blackmore2009convex, kariotoglou2013approximate}, as they over approximate failure probabilities. 
A known issue with these approaches is ``double counting''. A constraint violation of a trajectory at time $t$ might yield a violation at $t+1$ as well, and they will be considered as two separate violations because of the way they are constructed. In reality, we would like to consider that trajectory as just one failing trajectory.

Another common approach is to use \textbf{Monte-Carlo methods} to verify systems. These methods are very attractive because they can be applied to non-linear systems, intricate noise models and black-box simulators. However, they can be computationally inefficient, especially when trying to detect rare events~\cite{driveteslacanada:2021}. The verification process needs to iterate through many simulations or even guided simulations to find rare events~\cite{o2018scalable,verifai-cav19,schmerling2016evaluating} in order to produce an accurate estimate of the probability. Other more generic techniques also exist to improve the performance of a Monte-Carlo simulation~\cite{van2018introduction}.

An issue with Monte-Carlo simulations is when there are many non-deterministic parameters ~\cite{gessner2020integrals}; in such cases, Monte-Carlo techniques may require a prohibitively large number of simulations to adequately represent the posterior distribution. In our case, a dynamic system with multiple noise sources and a long time horizon can grow to a large parametric space quickly. We, and the work in \cite{gessner2020integrals} which we extend upon, show that our method can yield accurate integrations for Gaussians in high dimensions independent of the probability mass of the posterior.

\textbf{Optimization} techniques have been used extensively in the literature to verify systems with uncertainty. There exist numerous verification algorithms that deal with machine learning components in the loop (e.g.~\cite{tran2020verification}).
The authors in \cite{xiang2018reachability} consider dynamic systems with a neural network component as the controller. In that setting, the inputs to the neural network are discretized and a linear program over-approximates the output. With that, they compute the over-approximation of the complete system's reachable set.
Optimization can even help detect or attenuate cyber-physical attacks where an attacker can inject noise to a sensor to affect the system's outcomes through the controls~\cite{murguia2017reachable}. This assumes that bounded noises are selected carefully to achieve the attacker's goal while avoiding detection, and thus would be over-conservative in a non-adversarial setting. In \cite{dean2020robust}, the authors use robust control theory to provide guarantees for a system where the perception errors can be bounded using some assumptions on the data used during the training process versus the data that is collected in real-time.

Another line of work can be categorized as \textbf{geometric} algorithms. Set propagation techniques have been applied to reachability analysis. Except for a limited number of systems, these techniques always deal with under or over-approximations because finding the reachable sets is undecidable \cite{althoff2021set}. These techniques provide efficient computation frameworks; however, they work only on uni-model disturbances such as a Gaussian model, or a bounded disturbance. 
When discussing systems with a large number of states, one must employ other methods, such as decomposition of the system dynamics, for the methods to be tractable. The authors in \cite{althoff2016combining} combined zonotopes and support functions to create an efficient framework for calculating the reachable sets of linear and switched dynamics systems. It considers only bounded disturbances.

In this work, we focus on the verification of properties that can be expressed using Signal Temporal Logic (STL)~\cite{donze2010robust} formulas for linear or linearizable time-variant robotic systems with Gaussian error models. We show how our verification technique performs with a Gaussian mixture noise model (Section~\ref{sec:guassian_mix}), where the weights or probabilities of each Gaussian could be either static, come from a choice model like a Markov chain, or from a black-box choice model. %
We provide a verification method for generic STL formulae. We also describe a special case of reach-avoid~\cite{fan2018controller}  %
type specifications for which we propose an alternate solution that, in some cases, is more computationally efficient. We leverage and extend the framework in \cite{gessner2020integrals} to compute the probability that the robot satisfies or violates its task specification.

Our main \textbf{contribution} is a computation framework for verifying and computing the probability of a high-dimensional system to satisfy (or violate) complex STL specifications within a finite horizon using the STL quantitative semantics. The technique is especially useful (accurate and tractable) when dealing with low probability events and displays the following properties: 
1. We provide an efficient computation framework that does not suffer from the combinatorial nature of representing the signals that satisfy an STL specification. 
2. Failure modes are not double-counted and not over-approximated. The computational framework is solved efficiently and can be parallelized. 
3. Sampling is done from the posterior distribution in a rejection-free manner. Thus it is sampling efficiently from the target distribution.
4. We can efficiently sample \emph{new} trajectories from the failing or succeeding trajectory sets for analysis purposes, control synthesis, etc.
5. The algorithm is parameter-free. Meaning, no fine-tuning of hyper-parameters is required.
6. It can verify systems with more intricate noises than Gaussian errors thus capturing realistic perception models.

\section{Preliminaries}
\label{sec:Preliminaries}
In this section, we provide the necessary background on elliptical slice sampling and STL specifications.  

\subsection{Elliptical Slice Sampling (ESS) and the Holmes-Diaconis-Ross (HDR) algorithm}
\noindent An adaptive elliptical slicing method is used to sample from a linearly constrained domain under Gaussian distributions in \cite{gessner2020integrals}. We show the main concept and idea here for clarity and completeness. In this paper we extend \cite{gessner2020integrals} to compute the probability that the robot trajectories, represented as a multivariate Gaussian, satisfy or violate a specification.

\begin{figure}
     \centering
     \begin{subfigure}[b]{0.49\columnwidth}
         \centering
         \includegraphics[width=\columnwidth,height=3.5cm,keepaspectratio]{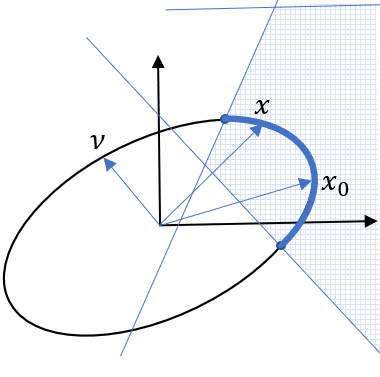}
         \caption{Elliptical slice sampling}
         \label{fig:ess_ell}
     \end{subfigure}
     \hfill
     \begin{subfigure}[b]{0.49\columnwidth}
         \centering
         \includegraphics[width=\columnwidth,height=3.5cm,keepaspectratio]{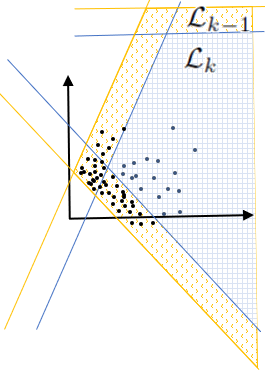}
         \caption{HDR}
         \label{fig:ess_hdr}
     \end{subfigure}
    \caption{(a) Sampling a new point $x\left(\theta^*\right)$ from the constrained domain (blue grid) given an initial point $x_0$, an auxiliary point $\nu \in \mathcal{N}(\mu, \Sigma)$ and $\theta^* \sim \mathcal{U}(\theta_{min}, \theta_{max})$. The active intersection is the bold blue line section of the ellipse where all points within $[\theta_{min}, \theta_{max}]$ are in the linearly constrained domain. (b) Original constrained domain $\mathcal{L}_K$ in the blue grid and shifted domain $\mathcal{L}_{K-1}$ in yellow divot after producing samples from $\mathcal{N}(\mu, \Sigma)$ using the ESS procedure under $\mathcal{L}_{K-1}$.}
    \label{fig:ess_and_hdr}
    \Description{In figure (a), an ellipse is formed using an initial sample x-naught and a randomly sampled nu from the original Gaussian distribution. The ellipse intersects with the domain of interest. The ellipse arc segment that lies over the domain is emphasized and is called the active domain. Another sample x is sampled with the ESS and is within the active domain as described in the text. Figure (b) shows the original domain of interest and the inflated domain. It also shows samples obtained via ESS in the inflated domain, and some of the samples also lie within the original domain. This ratio of samples is exactly the conditional probability of a single nesting.}
\end{figure}

Elliptical slice sampling (ESS) \cite{murray2010elliptical} is a Markov Chain Monte Carlo technique (MCMC) for sampling from a posterior when the prior is a multivariate Gaussian $\mathcal{N}(\mu, \Sigma)$. In our case, the posterior will be a Gaussian under constrained linear domains (a truncated Gaussian).
Given a single sample $x_0 \in \mathbb{R}^n$ inside the linear constrained domain $\mathcal{L}\subseteq \mathbb{R}^n$, and a new auxiliary point sampled from the same Gaussian $\nu \sim \mathcal{N}(\mu, \Sigma)$, the approach constructs an ellipse $x(\theta)=x_0 \cos(\theta)+\nu \sin(\theta)$, parameterized by the scalar $\theta \in [0,2\pi]$. Using a closed-form solution to the intersections between the auxiliary ellipse and the hyperplanes that confine the linear domain $\mathcal{L}$, we can sample $\theta^*$ from a Uniform distribution over the ellipse arc lengths that lie within the domain, and thus obtain a new sample $x(\theta^*) \in \mathcal{L}$. A point on the ellipse is in the domain $\mathcal{L}$, when the intersection between all $d$ constraints exceed zero, $Ax+b\ge 0$ where $A\in\mathbb{R}^{d\times n}, b\in\mathbb{R}^{d}$. This process is depicted in Fig.\ref{fig:ess_ell} where the new sample $x$ is sampled from the constrained Gaussian distribution $\mathcal{N}(\mu, \Sigma)$ (for proof, see \cite{murray2010elliptical,gessner2020integrals}).

The Holmes-Diaconis-Ross (HDR) algorithm~\cite{diaconis1995three} is one method from a family of algorithms called multi-level splitting, for estimating the probability of sampling from a constrained region under any distribution. Direct Monte-Carlo methods may be inefficient because most candidate samples may be rejected (low probability distribution function or high dimensional domain). With HDR, the probability $p(\mathcal{L})$ of sampling from $\mathcal{L} \subseteq \mathbb{R}^n$ is estimated using the product of conditional probabilities: 
\begin{equation}
 p(\mathcal{L})=p(\mathcal{L}_0) \prod_{k=1}^K p(\mathcal{L}_k | \mathcal{L}_{k-1})   
\end{equation} where $\mathcal{L}_0=\mathbb{R}^n,~p(\mathcal{L}_0)=1$. Each domain $\mathcal{L}_k$ (also referred to as a nesting) is shifted (enlarged, see Fig.\ref{fig:ess_hdr}) to $\mathcal{L}_{k-1}$ by a scalar $\gamma_k>0$ such that the conditional probabilities $p(\mathcal{L}_k | \mathcal{L}_{k-1}) \approx \frac{1}{2}$ and $\gamma_K=0$ is exactly the target domain $\mathcal{L}_K=\mathcal{L}$. 
Fig.\ref{fig:ess_hdr} depicts this process where the target domain $\mathcal{L}$ (blue grid) is expanded until it contains enough samples - when the probability to sample from the shifted region is about 0.5. Then, the algorithm iteratively shrinks the shifted region to keep the proportion of samples within the new domain to the previous domain at about half. $n_k$ samples are drawn from each domain $\mathcal{L}_{k-1}$ with the ESS algorithm. The probability $p(\mathcal{L}_k|\mathcal{L}_{k-1})=N(k)/n_k$, 
is the ratio between the number of samples $N(k)=\sum_{j=1}^{n_k}{I(x_j\in\mathcal{L}_k})$ ($I$ is the indicator function, equals one if the argument is true, zero otherwise) to the total number of samples, $n_k$, drawn at that nesting.

We note that closed-form solutions to the integral of a Gaussian under a linear constrained domain does not exist in the general case, when the domain is not axis-aligned with the Gaussian. Numerical methods, such as quadrature algorithms, do not scale well with the dimensionality of the problem \cite{orive2020cubature}.

\subsection{Signal Temporal Logic}

\noindent Signal temporal logic (STL) \cite{maler2004monitoring} enables specifying a broad range of temporal constraints over real-valued signals. 
Here we consider STL for discrete-time signals. Continuous-time logics and their properties can be found in, e.g., \cite{fainekos2009robustness,donze2010robust}. 

Consider a discrete-time real-valued signal ${\bf s} = s_0, s_1, s_2, \cdots$, where $s_t \in \mathbb{R}^n, \forall t \in \mathbb{N}$. A \emph{predicate} over $\mathbb{R}^n$ is denoted by $\mu=(h(s) \ge 0)$, where $h: \mathbb{R}^n \rightarrow \mathbb{R}$. A predicate is called \emph{linear} if $h$ is an affine function of $s$. Given a set of predicates, STL formulae are defined recursively using the following operators:
\begin{equation}
   \mu ~|~ \neg \varphi ~|~  \varphi_1 \wedge \varphi_2 ~|~  \varphi_1 \vee \varphi_2 ~|~
    \varphi_1 \mathcal{U}_{[t_1,t_2]} \varphi_2 ~|~  \Diamond_{[t_1,t_2]} \varphi ~|~ \Box_{[t_1,t_2]} \varphi
\end{equation}
where $\varphi, \varphi_1, \varphi_2$ are STL formulae, $\neg$ is the negation operator, $\wedge, \vee$ are conjunction and disjunction, respectively, and $\mathcal{U}_{[t_1,t_2]}, \Diamond_{[t_1,t_2]}, \Box_{[t_1,t_2]}$ are bounded temporal operators, over the time interval $[t_1,t_2]$, that stand for ``until'', ``eventually'', and ``always'', respectively. 
\begin{example}
\label{example_signal}
Consider a signal with values in $\mathbb{R}^2$, where $s=(s_{(1)},s_{(2)})'$. The specification
$$\varphi= \Box_{[0,9]} (s_{(1)}+s_{(2)}-10 \ge 0) \vee \Diamond_{[0,15]} \Box_{[0,5]} (-s_{(1)} \ge 0)$$
encodes ``for all times in the interval [0,9], the value of $s_{(1)}+s_{(2)}$ stays above $10$, \emph{or}, for some time in the interval $[0,15]$, the value of $s_{(1)}$ stays below $0$ for $5$ consecutive time steps".
\end{example}

\begin{define}
The STL score, or quantitative semantics~\cite{donze2013}, $\rho({\bf s},\varphi,t)$ is recursively defined as:
\begin{itemize}[leftmargin=*]
    \item $\rho({\bf s},(h(s)\ge 0),t)=h(s_t)$,
    \item $\rho({\bf s},\neg \varphi,t)=-\rho({\bf s},\varphi,t)$,
    \item $\rho({\bf s},\varphi_1 \wedge \varphi_2,t)=\min(\rho({\bf s},\varphi_1,t),\rho({\bf s},\varphi_2,t))$, 
        \item $\rho({\bf s},\varphi_1 \vee \varphi_2,t)=\max(\rho({\bf s},\varphi_1,t),\rho({\bf s},\varphi_2,t))$, 
    \item  $\rho({\bf s},\Diamond_{[t_1,t_2]} \varphi, t) = \underset{\tau \in t+[t_1,t_2]}{\max} \rho({\bf s},\varphi,\tau)$, 
     \item  $\rho({\bf s},\Box_{[t_1,t_2]} \varphi, t) = \underset{\tau \in t+[t_1,t_2]}{\min} \rho({\bf s},\varphi,\tau)$,
        \item $\rho({\bf s},\varphi_1 \mathcal{U}_{[t_1,t_2]} \varphi_2, t) = \underset{\tau \in t+[t_1,t_2]}{\max} \min\big(\rho({\bf s},\varphi_2,\tau),\underset{\tau' \in [t,\tau]}{\min}\rho({\bf s},\varphi_1,\tau')\big)$. 
\end{itemize}
\label{define_stl_score}
\end{define}

The STL score provides a metric for distance to satisfaction for a signal and STL formula. A positive STL score indicates satisfaction and a negative one stands for violation. To remove ambiguity, we consider the STL score of $\rho(s,\varphi,t)=0$ as satisfying. We define the STL score of a signal ${\bf s}$ and specification $\varphi$ as $\rho({\bf s},\varphi,0)$.  

\begin{example}
In Example \ref{example_signal}, let $s_t=(t-8,2)',  \forall t \in \mathbb{N}$. It is evident that it satisfies $\varphi$. Applying Definition \ref{define_stl_score}, we obtain
$$\rho({\bf s},\Box_{[0,9]} (s_{(1)}+s_{(2)}-10 \ge 0),0)=\underset{t \in [0,9]}\min (t-16)=-16$$
and 
$$\rho({\bf s},\Diamond_{[0,15]} \Box_{[0,5]} (-s_{(1)} \ge 0) ,0) = 
\max_{t \in [0,15]} \min_{\tau' \in [0,5]}(8-(t+\tau'))=3,
$$
thus $\rho({\bf s},\varphi,0)= \max(-16,3)=3$. We say signal ${\bf s}$ satisfies $\varphi$ and its STL score is 3. 
\end{example}
\begin{define} The $\varrho$-level set of an STL formula $\varphi$ is defined as:
\begin{equation}
    \mathcal{L}(\varphi,\varrho) = \{ {\bf s} | \rho({\bf s}, \varphi, 0) \ge  \varrho \}.
\end{equation}
\end{define}
\begin{define}
The horizon of the STL formula $\varphi$, denoted by $H^\varphi$, is the minimum length of truncated signal ${\bf s}=s_0,s_1,\cdots,s_{H^\varphi-1}$ such that is required to evaluate $\rho({\bf s},\varphi,0)$ and it is recursively given by:
\begin{itemize}[leftmargin=*]
    \item $H^{(h(s)\ge 0)}=1$,
    \item $H^{\neg \varphi}=H^{\varphi}$,
    \item $H^{\varphi_1 \wedge \varphi_2}=H^{\varphi_1 \vee \varphi_2}=\max(H^{\varphi_1}, H^{\varphi_2}) $,
    \item $H^{\Diamond_{[t_1,t_2]} \varphi} = H^{\Box_{[t_1,t_2]} \varphi} = t_2+   H^{\varphi} $
        \item $H^{\varphi_1 \mathcal{U}_{[t_1,t_2]} \varphi_2} =  t_2 + \max(H^{\varphi_1}, H^{\varphi_2})$.
\end{itemize}
\end{define}
\begin{example}
In Example \ref{example_signal}, the horizon of the formula is $H^\varphi=\max(9+1, 15+5+1)=21$. The values of $s_{21},s_{22}, \cdots $ do not affect $\rho({\bf s},\varphi,0)$.  
\end{example}
Given $\varphi$, we only need the truncated signal $s_0,s_1,\cdots,s_{H^\varphi-1}$ to check whether it satisfies $\varphi$. Thus, we can stack the truncated signal into a vector denoted by $s^\varphi:=(s_0',s_1',\cdots,s_{H^\varphi-1}')' \in \mathbb{R}^{n\cdot H^\varphi}$. 
With a slight abuse of notation we extend the STL score and level-set definitions to the following function and set in $\mathbb{R}^{n\cdot H^\varphi}$:
\begin{equation}
    \rho(s^\varphi):=\rho({\bf s}, \varphi, 0).
\end{equation}
\begin{equation}
    \mathcal{L}(\varphi,\varrho):=\{ s^\varphi \in \mathbb{R}^{n\cdot H^\varphi} | \rho(s^\varphi)\ge \varrho \}.
\end{equation}
It is straightforward to show that given $\varphi$ with linear predicates on $\mathbb{R}^n$, we have the following properties:
\begin{itemize}
\item The function $\rho : \mathbb{R}^{n\cdot H^\varphi} \rightarrow \mathbb{R}$ is piecewise affine and Lipschitz continuous. 
\item For a given $\varrho$, the set $\mathcal{L}(\varphi,\varrho)$ is a union of polyhedra in $\mathbb{R}^{n\cdot H^\varphi}$. 
\end{itemize}

\section{Problem setup}
\label{sec:problem_setup}

\subsection{System}
\label{sec:prob_system}

\noindent We consider discrete linear(izable), possibly time-varying, systems (LTV) with the dynamic and measurement equations: 
\begin{align}
\label{eqn:sys_dyn}
    x_{t+1}&=A_t x_t + B_t u_t + w_t, \\\nonumber
    y_{t}&=C_t x_t + v_t%
\end{align}
where $x_t \in \mathbb{R}^n$ is the state at time $t$, $u_t \in \mathbb{R}^m$ is the control input and $y_t \in \mathbb{R}^q$ is the measurement vector. The process noise $w_t \in \mathbb{R}^n$ and the measurement noise $v_t \in \mathbb{R}^q$ are described in more detail in  Section~\ref{sec:prob_perception}. $A_t, B_t$ and $C_t$ are the relationships between the states and measurements and are assumed known. The discrete system has a time step size of $\Delta t$. The system can have a linear state observer, and a closed loop feedback controller for tracking a reference trajectory $r_t$:
\begin{equation}
\label{eqn:sys_obs}
    \hat{x}_{t+1}=A_t \hat{x}_t + B_t u_t + L_t (y_t-C_t \hat{x}_t)
\end{equation}
\begin{equation}
\label{eqn:sys_ctrl1}
    u_{t}=r_t - K_t \hat{x}_t
\end{equation}
or, directly using the measurement for feedback: \begin{equation}
\label{eqn:sys_ctrl2}
    u_{t}=r_t - K_t y_t.
\end{equation}

\subsection{Noise model}
\label{sec:prob_perception}

\noindent In this paper we focus on Gaussian errors, $v_t \textapprox  \mathcal{N}(\mu_t^v, \Sigma_t^v)$ and $w_t \textapprox \allowbreak  \mathcal{N}(\mu_t^w, \Sigma_t^w)$. We assume that all the noises are independent and identically distributed (iid). Note that one can augment the system's states if the noise is colored. 

\sloppy In Sec.~\ref{sec:guassian_mix} we discuss a more intricate noise model, where the noise is modelled as a Gaussian mixture, meaning $v_t  \textapprox \sum_{m=1}^{M_v}{ \pi_m^v \mathcal{N}(\mu_m^v, \Sigma_m^v) }$ where $\pi_m^v$ is the probability of choosing Gaussian distribution $m$ (similarly for $w_t$). While a single Gaussian is a special case of the mixture, we separate the discussion because we can provide stricter guarantees for this case. %

\subsection{Specification}
\label{sec:prob_specs}

We consider STL specifications where the underlying signal is the system trajectories:
\begin{equation}
    {\bf x}=x_0, x_1, \cdots.
\end{equation}
We limit ourselves to linear predicates on the system's state in the form of $\mu=(a'x+b \ge 0), a \in \mathbb{R}^n, b \in \mathbb{R}$. The assumption of linearity is essential since later in the paper we will use a closed-form solution for intersections of an ellipse and a hyperplane (Section \ref{sec:ess_hdr_step}). While it is possible to consider specific forms of nonlinear predicates and still retain closed-form solutions, we leave that to future work.   

\begin{example}
\label{ex:reach_avoid}
A common STL formula is \emph{reach-avoid}. Consider a continuous-time system with a time horizon $T$  with an STL specification of this structure:
\begin{equation}
\label{eqn:prob_spec}
\varphi_{R/A}  :=  \phi_0 \wedge \bigwedge_{i=1}^{N_\text{unsafe}} \Box_{\left[0,T\right]} \neg \phi_{\text{unsafe},i} \wedge \bigwedge_{j=1}^{N_\text{goals}} \Diamond_{[T_{0_j},T_{1_j}]} \phi_{\text{goal},j},
\end{equation}
where $T\ge T_{1_j}, j=1,\cdots,N_\text{goals}$ and $\phi_0 = \bigwedge_{i=1}^{N_0} (a'^0_i x + b^0_i \ge 0)$ defines a polyhedron in the state-space with $N_0$ hyperplanes each represented by a linear predicate. Similar notation is used to define sets of polyhedra for the unsafe sets (e.g. obstacles) and the goals.  
In words, the system satisfies the specification when it is able to start in the set defined by $\operatorname{Init}$, avoid all obstacles $\operatorname{Unsafe}$ for the entire trajectory and reach each $\operatorname{Goal}_j$ at some $t \in [T_{0_j},T_{1_j}]$. Given $\Delta t$ and $T$, a trajectory of the system contains $t_H=\lceil T/\Delta t\rceil$ discrete time steps. To be able to correctly verify the specification $\varphi$, we require $t_H \geq H^\varphi$. %
\end{example}%

An advantage of our approach is the ability to efficiently address the combinatorial aspect of all possible trajectory classes that may satisfy or violate the specification without double counting them.

\subsection{Problem formulation}
\label{sec:prob_form}

\begin{problem}
\label{problem}
Given a linear system in the form of \eqref{eqn:sys_dyn}-\eqref{eqn:sys_ctrl2}, a Gaussian (mixture) noise model and an STL formula $\varphi$ with linear predicates over $x$, find the probability that $\varphi$ is satisfied. 
\end{problem}

\section{Approach}
\label{sec:approach}

We illustrate our approach through  an example of a holonomic robot navigating in a workspace (Fig.~\ref{fig:roll_example_config}).

\begin{example}
\label{example_holonomic_robot}
A holonomic robot's state is $\xi=[x,y,\dot{x},\dot{y}]^\prime$ with the discrete-time dynamics:
\begin{flalign}
\label{eqn:holonom_dyn_approach}
\xi_{t+1}=&
\begin{bmatrix}
    1 & 0 & \Delta t & 0 \\
    0 & 1 & 0 & \Delta t \\
    0 & 0 & 1 & 0 \\
    0 & 0 & 0 & 1 
\end{bmatrix} \xi_t+ 
\begin{bmatrix}
    \sfrac{\Delta t^2}{2m} & 0 \\
    0 & \sfrac{\Delta t^2}{2m} \\
    \sfrac{\Delta t}{m} & 0 \\
    0 & \sfrac{\Delta t}{m} 
\end{bmatrix} u_t + w_t \\\nonumber
u_t=&r_t - 
K_{fb} \eta_t
\end{flalign}
We use the discrete Linear Quadratic Regulator (LQR) algorithm \cite{kwakernaak1972linear} with the desired $\Delta t$ to compute the optimal controller $K_{fb}$. We assume full-state measurement:
\begin{flalign}
\label{eqn:holonom_meas_approach}
\eta_{t}= \xi_t+v_t
\end{flalign}
The noise $v_t$ is normally distributed and $w_t$ is omitted for brevity. We consider an arbitrary STL specification $\varphi$ %
for the rest of the section unless otherwise specified.
\end{example}

\subsection{Integral over Trajectory Space}
\label{sec:traj_space}
The first step is to derive the relationship between the process and measurement noises to the trajectories by turning the trajectories into Gaussians in a higher dimensional space $x_{traj}\triangleq[x_0',\dots,x_{t_H-1}']'\in\mathbb{R}^{n \cdot t_H}$. %
 In our robot example, this means the concatenated states for every time step in the horizon, ${\bf x}_{traj}\in\mathbb{R}^{4\cdot t_H}$. We consider $w_t=0$ and $A_t=A, B_t=B$ and $C_t=C$ without loss of generality to simplify the following expressions. Based on Eq. \eqref{eqn:sys_dyn}-\eqref{eqn:sys_ctrl2}, we can express the full trajectory in vector form with \eqref{eqn:sys_resp} by iteratively substituting the states and controls:
\begin{flalign}
\label{eqn:sys_resp}
   &{\bf x}_{traj} =
    \Phi_0 x_0 + \Phi_r {\bf R} + \Phi_v {\bf V}
\end{flalign}

\noindent Where ${\bf R}=[r_0',\dots,r_{t_H-1}']'\in\mathbb{R}^{m\cdot t_H}$, and ${\bf V}=[v_0',\dots,v_{t_H-1}']'\in\mathbb{R}^{q\cdot t_H}$. $\Phi_0,\Phi_r,\Phi_v$ are the matrix coefficients that transfer the initial state, the extended reference inputs and measurement noises to the full trajectory, respectively. All components in \eqref{eqn:sys_resp} are deterministic except for the stochastic ${\bf V}$ according to the sampled noise mode:
\begin{align}
\label{eqn:noise_distribution}
 V \sim \mathcal{N}(
    [\mu_0^{v'},\cdots,\mu_{t_H-1}^{v'}]',diag([\Sigma_0^v,\cdots,\Sigma_{t_H-1}^v])
\end{align}
We can extract the multivariate Gaussian in the trajectory space which is the distribution over which we integrate:
\begin{flalign}
\label{eqn:trajectory_pdf}
    {\bf x}_{traj} \sim \mathcal{N}\big(& \Phi_0 x_0 + \Phi_r {\bf R}+ 
    \Phi_v \operatorname{M}, 
        \Phi_v 
        \Sigma \Phi_v^\prime \big)
\end{flalign}
$M=[\mu_0^{v'},\cdots,\mu_{t_H-1}^{v'}]'$. In a similar manner, it is possible to derive the Gaussian of a trajectory with both $v_t$ and $w_t$ (and possibly, $x_0 \sim \mathcal{N}$). 

Other work, e.g. chance constraints that are typically implemented and over-approximated with Boole's inequality, deal with constraints on the state-level. We work with the full trajectory. This difference is one of the reasons we do not double count events. When computing the probability of failures, the trajectory Gaussian is integrated with respect to the trajectory-level constraints.

The evaluation of the probability in Problem \ref{problem} is equivalent to computing the following integral:
\begin{equation}
\label{eq_integral}
    p(\varphi)= \underset{x^{\varphi} \in \mathcal{L}(\varphi,0)}{\int} \operatorname{pdf}(x^{\varphi}) d{x^{\varphi}},
\end{equation}
where $\operatorname{pdf}(x^{\varphi})$ is the probability density function of the trajectories, the Gaussian in this case from \eqref{eqn:trajectory_pdf}. $\mathcal{L}(\varphi,0)$ is the set where the trajectories satisfy the specification $\varphi$.

\subsection{Monte-Carlo Sampling}
\label{sec:approach_overview}
\noindent We propose a guided Monte-Carlo approach for verifying a dynamical system with fixed controls (they can be time-varying but not state-dependent) and Gaussian noise sources over a fixed horizon with respect to STL specification. %
We represent the full trajectories as Gaussian, and use the HDR and ESS algorithms to integrate the probability density function under the domains that satisfy the STL formula. %

The advantages of the approach are threefold: 1. Efficient (rejection-free and parameter-free) sampling of trajectories that satisfy a STL specification and computation of the probability of satisfaction, without over-approximations and double-counting, such as with the use of Boole's inequality on each separate state in the trajectory. 2. Efficiently finding events with low probability that would be otherwise intractable to compute with naive Monte-Carlo simulations. 3. It enables longer horizons and more random variables without suffering from the dimension explosion problem (the ill-sampling of the posterior distribution).

The first point is achieved by sampling, with ESS, trajectories that are within the set of trajectories that satisfy the specification. The second point is achieved using the HDR algorithm as we can construct the required number of nestings to evaluate the probability. In fact, once all nestings are set up, the sampling time of the rejection-free ESS algorithm is not influenced by the probability mass.
Regarding the third point, the variance of the error of the quantity we wish to estimate with a Monte-Carlo simulation $\sigma^2_{\bar{x}}=\sigma^2_{x}/n_{sim}$ decreases with the number of simulations $n_{sim}$. However, we cannot accurately estimate the value of $\sigma^2_x$ from the sampled simulations when we ill-sample the posterior distribution. We do not know the true variance a priori and in fact, the variance itself may increase rapidly as the number of variables increase. Intuitively, there are more combinations of noise errors which may cause the robot to violate the specification and it is harder to sample ``useful'' (for the purpose of correctly estimating the probability) combinations. Our approach, on the other hand, is sampling rejection-free from the constrained posterior distribution to the requisite level of accuracy.

\begin{figure}
     \centering
     \includegraphics[width=\columnwidth,height=5cm,keepaspectratio]{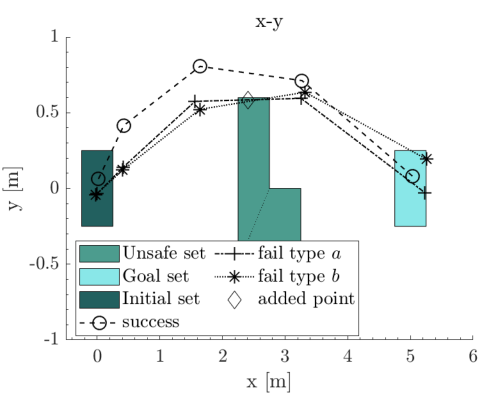}
     \caption{Example of a robot in 2-D. The initial, unsafe and goal sets are depicted in the figure. Successful and failing trajectories of a $5$-step horizon are shown with an added intermediate point. We define failure trajectories that hit an obstacle, but reach the goal on time as type $a$;  type $b$ failing trajectories that do not reach the goal on time. In Section \ref{sec:coll_avoid} we make use of these failures for reach-avoid specifications.} %
     \label{fig:roll_example_config}
     \Description{Figure shows an x-y projection for a 2-D navigating robot example. Robot goes from initial set on the left, to the goal set on the right while avoiding an L shaped obstacle in the middle. Three trajectories show - a successful example; an example that fails due to hitting an obstacle (type a); and an example that does not reach the goal in time (type b). Also, a marker that shows an added point between two time steps to increase computation fidelity.}
\end{figure}

\subsection{STL-Score-Guided Elliptical Slice Sampling}
\label{sec:ESS_STL}
Here we describe how we draw sample points from $\text{pdf}(x)$ that are inside $\mathcal{L}(\varphi,\varrho)$ - trajectories that have an STL score $\ge \varrho$. As mentioned earlier, the naive way is to draw samples from $\text{pdf}(x)$ and reject those that fall outside of $\mathcal{L}(\varphi,\varrho)$. However, if the probability mass inside $\mathcal{L}(\varphi,\varrho)$ is too small, the procedure will be inefficient as most of the samples will be rejected.

We use ESS as described in Section \ref{sec:Preliminaries}. The explicit representation of $\mathcal{L}(\varphi,0)$ - the domain of the integral in \eqref{eq_integral} - as a union of polyhedra requires an enumeration of all of the possible convex sets. The number of such sets can grow exponentially in the size of the formula (see, e.g., \cite{sadraddini2016feasibility}). We avoid explicit enumeration of the polyhedra in $\mathcal{L}(\varphi,0)$ while computing the integral in \eqref{eq_integral}. The key insight is that we only need the STL score function \cite{nivckovic2020rtamt}.

\begin{theorem}
\label{tm1}
Given a STL formula $\varphi$ with a set of linear predicates $\mu_i=(a'_ix+b_i \ge 0), i=1,\cdots,N_\varphi$, where $N_\varphi$ is the total number of predicates. Given an existing sample trajectory $x_{e}^\varphi \in \mathcal{L}(\varphi,\varrho)$ and free sample trajectory $x_{f}^\varphi$ (not necessarily in $\mathcal{L}(\varphi,\varrho)$), construct the ellipse $\mathcal{E} = \{x_{e}^\varphi \cos \theta+  x_{f}^\varphi \sin \theta~|~\theta \in [0,2\pi]\}$ in $\mathbb{R}^{n.H^\varphi}$. Then construct the following sorted list of real numbers in $[0,2\pi]$:
\begin{equation}
\label{eq_Theta}
\begin{array}{ll}
\Theta = \text{sorted} \big \{  &  \theta~|~\exists t \in \{0,1,\cdots,H^\varphi-1\}, i \in \{1,\cdots,N_\varphi\}, \\
& \text{ s.t. }a'_i x_t+b_i = \pm \varrho, \\ & x^\varphi = x_e^\varphi \cos \theta+  x_f^\varphi \sin \theta, \\
& x^\varphi=(x_0',x_1',\cdots,x_{H^\varphi}')'\big \}.
\end{array}
\end{equation} 
Then for any two consecutive elements $\theta_1,\theta_2 \in \Theta$ (cyclic), one of the following statements is correct:
\begin{equation}
\forall \theta \in [\theta_1,\theta_2], \rho(x_e^\varphi \cos \theta+  x_f^\varphi \sin \theta) \ge \varrho, \text{ or}
\end{equation}
\begin{equation}
\forall \theta \in [\theta_1,\theta_2], \rho(x_e^\varphi \cos \theta+  x_f^\varphi \sin \theta) \le \varrho.
\end{equation}
\end{theorem}
\proof{
In order for $\rho(x^\varphi)=\varrho$, the value inside the function of at least one of the predicates should be equal to $\pm \varrho$ - this predicate becomes the maximizer/minimizer in the STL score function. Note that we have $\pm$ as negation might be in the formula. Therefore, the set $\Theta$ contains all the roots for $\rho(x^\varphi) - \varrho = 0$ - but can contain spurious elements. Since $\rho$ is Lipschitz continuous, $\rho(x^\varphi) - \varrho$ is sign-stable on $\mathcal{E}$ between two consecutive roots. 
}

\begin{figure}
     \centering
     \includegraphics[width=\columnwidth,height=4cm,keepaspectratio]{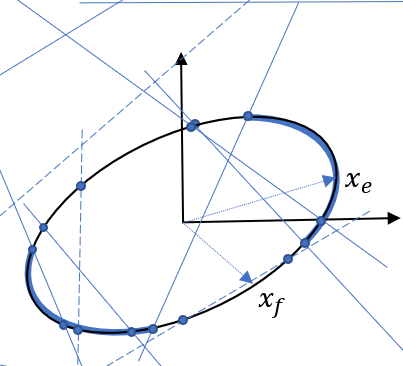}
     \caption{Hyperplanes, representing predicates in the STL formula, projected on the auxiliary ellipse. Some may not intersect, and some may intersect but not change $\mathcal{L}(\varphi,\varrho)$ thus are in the domain. Dashed lines represent hyperplanes of predicates that are not in the time bounds described in $\varphi$.}
     \label{fig:ess_fullstl}
     \Description{The ESS auxiliary ellipse intersected by multiple lines representing the hyperplanes of each predicate in the specification. Dashed lines represent hyperplanes that result from smearing the predicates in time and that will not play a role in the specification itself, because it is out of the specification's time bounds. The arc that points x1, x2 form on the ellipse are not intersected by any hyperplane, thus maintain the same sign of the robustness. The arc formed by x3,x4 is intersected with a dashed line which also means the sign of the robustness does not change between those points.}
\end{figure}

Theorem \ref{tm1} paves our way to compute portions of the ellipse that fall into $\mathcal{L}(\varphi,0)$ by only computing the roots of the robustness function on the ellipse. Furthermore, \eqref{eq_Theta}, provides all the candidates with the complexity of solving $2\cdot H^\varphi N_\varphi$ intersections of the ellipse with a hyperplane, for which closed-form solutions exist \cite{gessner2020integrals}. Then, using Theorem \ref{tm1}, we can sample $\theta \in [\theta_1,\theta_2]$ within each pair, to assign if $[\theta_1,\theta_2]$ is in $\mathcal{L}(\varphi,0)$.  See Fig.~\ref{fig:ess_fullstl} for an illustration of the adjusted ESS procedure.
Thus, the complexity of obtaining the intersection of the ellipse and $\mathcal{L}(\varphi,\varrho)$ is $\mathcal{O}(H^\varphi N_\varphi)$, and thus we avoided any exponential blow up due to explicit combinatorial representation of $\mathcal{L}(\varphi,\varrho)$.

\subsection{Holmes-Diaconis-Ross for STL}
\label{sec:ess_hdr_step}
Now that we have a method to draw samples from $\mathcal{L}(\varphi,\varrho)$, we use it for our HDR-based Monte-Carlo method.

\subsubsection{Nesting partitioning}
\noindent To perform HDR where the probability density function is low, we need to account for the multi-level splitting described in the preliminaries. This means that when sampling from the nesting $k$, a larger domain than what we would like to evaluate, we shift $\varrho$ to a new value (usually it will be a negative value, allowing more trajectories that violate $\varphi$, where about half the trajectories have robustness greater than $\varrho$) as the new cutoff level instead of $0$. On the other hand, we also need to shift the linear predicates, to get the new intersections of $\mathcal{L}(\varphi,\varrho)$. For a general specification, we do not know whether to shift the predicates with a positive or a negative $\varrho$ due to the structure of the sub-formulas. For example, consider the difference between $\varphi_1:=\operatorname{h}(s)$ versus $\varphi_2:=\neg \operatorname{h}(s)$. In $\varphi_1$, a violating sample would require increasing the domain, while a violating sample on $\varphi_2$ would decrease the domain to make it satisfying. Instead of analyzing each component of the specification, we  shift each predicate by $+\varrho$ and by $-\varrho$ as shown in \eqref{eq_Theta}. %

\subsubsection{Error Analysis}
\noindent Monte-Carlo methods by nature give different results every time they are executed. It is necessary to have an estimate on the variance of the computed probability $p(\varphi)$. For the HDR nesting $k$, we sample $n_k$ samples, and as discussed in Section~\ref{sec:Preliminaries}, we aim for the conditional probability to be $p_{k\given k-1} \approx 0.5$. In principal, ESS is a MCMC method, thus the samples are by definition dependent and the central limit theorem (CLT) does not apply. To mitigate this limitation, we keep only every $n_d$-th sample from the ESS, thus making the dependency between the sampled $x_i$ to $x_{i+n_d}$ practically non-existent (in all our examples we used $n_d=4$). This is sometimes referred to as the ``burn-in'' phase, and its purpose here is to weaken the dependency between samples. In practice, our experience shows (see Fig.~\ref{fig:mc_of_hdr_and_mc}) that the dependency is weak, due to the ``burn-in'' process and sampling $\theta$ independently from a uniform distribution, and the following error analysis applies.

Using the CLT ($n_k p_{k|k-1} \gg 1$), we can assess that the variance for nesting $k$ is $\sigma_k^2 \approx p_{k|k-1}(1-p_{k|k-1})/n_k$ where $p_{k|k-1}=\operatorname{N}(k)/n_k$ and $\operatorname{N}(k)$ is the number of points sampled within $\mathcal{L}_{k}$. Therefore, for nestings $k=\{1,\dots,K-1\}$, a good approximation for the variance is $\sigma_k^2 \approx \tfrac{1}{4 n_k}$, such that $p_{k|k-1}  \textapprox \mathcal{N}(\tfrac{1}{2}, \tfrac{1}{4n_k})$. %
With a slight abuse of notation, we define $p_{k}\equiv p_{k\given k-1}$ for clarity. Each $p_k$ is a Gaussian iid, we can compute the variance of the product of the conditionals:
\begin{flalign}
\label{eqn:var_of_nestings}
\operatorname{Var}[p_1\cdots p_K] 
&= \EX[(p_1\cdots p_K)^2]-\left(\EX[p_1\cdots p_K]\right)^2\\\nonumber
&= \EX[p_1^2\cdots p_K^2]-\left(\EX[p_1]\cdots \EX[p_K]\right)^2\\\nonumber
&= \EX[p_1^2]\cdots \EX[p_K^2] - (\EX[p_1])^2\cdots (\EX[p_K])^2\\\nonumber
&= \prod_{k=1}^K \left(\operatorname{Var}[p_k]+(\EX[p_k])^2\right)
- \prod_{k=1}^K \left(\EX[p_k]\right)^2
\end{flalign}
Substituting for our nominal parameters we obtain the approximation:
\begin{flalign}
\label{eqn:hdr_approx_var}
\operatorname{Var}[p_1\cdots p_K] = \prod_{k=1}^K \left(\frac{1}{4n_k}+\frac{1}{4}\right)
- \prod_{k=1}^K \left(\frac{1}{4}\right)
\end{flalign}
In practice, %
we compute the variance with the actual sampled values and not \eqref{eqn:hdr_approx_var}. The point is that when the number of points in a nesting is large enough, the variance is proportional to $(1/4)^{K-1}$. For example, selecting $64$ points per nesting with $K=5$ yields a standard deviation $\sigma=0.031$.

\subsubsection{Adaptive nesting samples}
\label{sec:adapt_num_samp}

\begin{figure}
     \begin{subfigure}[b]{0.6\columnwidth}
     \centering
     \includegraphics[width=\columnwidth,height=5cm,keepaspectratio]{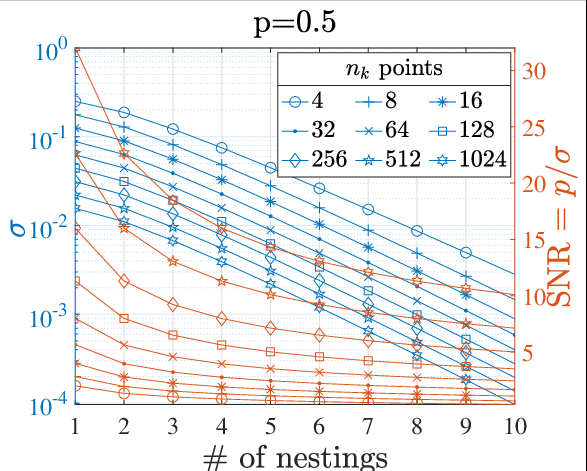}
     \caption{}
     \label{fig:hdr_adapt_samples}
     \end{subfigure}
     \hfill
     \begin{subfigure}[b]{0.39\columnwidth}
     \centering
     \includegraphics[width=\columnwidth,height=3.7cm]{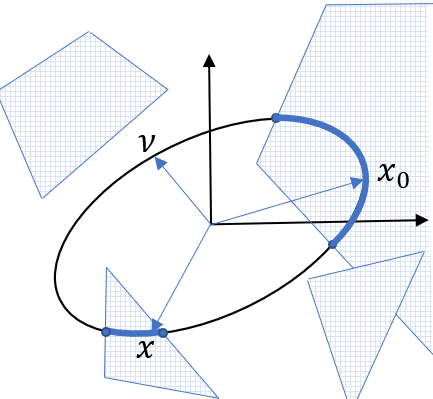}
     \caption{}
     \label{fig:ess_ell_union}
     \end{subfigure}
     \caption{(a) The effect of the number of nestings and the number of sampled points on the confidence interval of the HDR algorithm on a nominal $p_k=0.5$. (b) Extending \cite{gessner2020integrals} to  a union of polytopes.}
     \label{fig:adapt_nest_and_ess_union}
     \Description{}
\end{figure}

\noindent Using \eqref{eqn:var_of_nestings}, we can compute an expected minimal number of samples for a desired value of $p(\mathcal{L}_k | \mathcal{L}_{k-1})$. Fig.\ref{fig:hdr_adapt_samples} shows how increasing the number of samples decreases the uncertainty. Increasing the number of nestings, decreases the uncertainty as well. However, the number of nestings is not a design parameter but rather depends on the problem at hand. The number of nestings is approximately $K=\lceil-\operatorname{log}_2 p\rceil$. We can automatically select the number of samples to use per nesting by utilizing \eqref{eqn:var_of_nestings} and Fig.\ref{fig:hdr_adapt_samples}. The benefit of using this is shorter computation times;  in problems with a high number of nestings (i.e. low probability), we can get the desired confidence interval with less samples.

\subsection{Special case: Reach-avoid specifications}
\label{sec:coll_avoid}

\noindent In Eq. \eqref{eqn:prob_spec} we showed a common specification for robotic applications. In {some} cases, it might be more efficient to construct the union of polyhydra that represent the predicates in $\varphi_{R/A}$ 
explicitly, rather than use the STL score-based ESS and HDR (due to the number of hyperplanes and STL score computations discussed in Section \ref{sec:ESS_STL}).
Consider finding the probability of violating Example \ref{ex:reach_avoid}, i.e.  $\mathcal{L}( \neg\varphi_{R/A},0)$. We define two possible ways of violating $\varphi_{R/A}$ (i) type $a$:  $\varphi_a:=\phi_0 \wedge \Diamond_{[0,T]} \phi_{\text{unsafe}} \wedge \Diamond_{[T_{0},T_{1}]} \phi_{\text{goal}}$, ``hit an obstacle and reach the goals on time''. (ii) type $b$:  $\varphi_b:=\phi_0 \wedge \Box_{[T_{0},T_{1}]} \neg\phi_{\text{goal}}$, ``do not reach the goals on time''. Thus, $\mathcal{L}(\varphi_a,0) \bigcup \mathcal{L}(\varphi_b,0) =\mathcal{L}(\neg\varphi_{R/A},0)$. 

We compute the integral of the Gaussian under the constrained domain by modifying the procedure in \cite{gessner2020integrals}; we consider a union of polytopes instead of only one. This means that as long as $\exists l \in \operatorname{Set}(poly.)$ such that the intersection of all its constraints exceed zero, %
then it is a valid point in the domain. There may be numerous intersections of the constructed ellipse with the faces of the different polytopes. We extend \cite{gessner2020integrals} (Fig.~\ref{fig:ess_ell_union}) to find the active segments of the union of polytopes and sample points from the active domain.

We consider discrete time systems; however, there is a gap when verifying the system that is in fact continuous. There could be situations where all the discrete states in the horizon are satisfying the specification, yet the system might collide with an obstacle in between the states. See Fig.~\ref{fig:roll_example_config} for an example near the obstacle. 

There are several ways to address this. First, we can either increase the sampling rate or bloat the obstacles. The former will increase the dimension of the problem, while the latter would constrain the problem even more. In both cases, it will increase the computational load by increasing the dimensions in the trajectory space or by reducing the probability mass function under the domains (need more nestings).

The second approach, if we assume constant velocity between two consecutive states (valid in short time spans), we can introduce more constraints without increasing the problem dimensions. These intermediate points will add robustness by adding more area of the polytopes without as many computations as increasing the number of states. 
It can also be introduced only in parts of the trajectory that are susceptible to failure. Fig.~\ref{fig:roll_example_config} shows an example of a point added in the middle between $t_2$ to $t_3$. To add constraints for an intermediate centroid point between two trajectory points:
$$
\label{eqn:constr_collision}
a'_i(0.5x_t+0.5x_{t+1}) + b_i \ge 0
$$
This additive technique can also be applied to compute the $\rho(x^\varphi)$ if the STL library can compute the score of signals with dense time steps.

\subsection{Gaussian Mixture models}
\label{sec:guassian_mix}

\noindent Our approach may also be used to verify systems where the underlying noise model is better described with a Gaussian mixture. For example, a common model for range finders is the Beam model \cite{thrun2002probabilistic} (Ch. 6). It incorporates several modes of sensing errors that depend on the physical interaction of the sensor with its environment and may be approximated by a Gaussian mixture. Another example is a camera that is tracking cars but due to occlusions or errors in its neural net, it starts to track clutter or a different car in its field of view. The noise distribution at time $t$ might depend on the distribution at $t-1$, making $\pi_m^v$ come from a Markov chain or a black-box choice model: 
\begin{flalign}
\label{eqn:gauss_multi_dist}
v_t  \textapprox \sum_{m=1}^{M}{ \pi_m^v \mathcal{N}(\mu_m^v, \Sigma_m^v) }
\end{flalign}

\noindent Where $M$ is the number of distributions and can also vary between time steps.
In this case, computing the tree of possible combinations of the Gaussian distributions and noises throughout the trajectory and their weights is intractable. However, we suggest a procedure for computing the total probability. %
The first step samples just the mixing factors $\pi_m$ from their underlying distributions, for the entire horizon. When the mixing factors are fixed, the problem reduces to Problem \ref{problem}. We compute the probability $p(\varphi)$ and variance, and repeat this procedure for $N$ iterations. 
Then, we compute the unbiased mean estimate and the variance of the $N$ iterations. This method still relies on Monte-Carlo simulation to compute the probability and variance estimation. However, only the trajectory modes are sampled, thus reducing the problem's input dimensions considerably. A full Monte-Carlo simulation will have the modes and the actual values to sample from and can thus be susceptible to the dimension explosion problem.

\section{Case studies}
\label{sec:demos}

\subsection{Robot navigation - reach-avoid}
\label{sec:demo_robot_nav}

\noindent We demonstrate verification for Example \ref{example_holonomic_robot}. 
The noises are: $v_x \textapprox \mathcal{N}(0,0.06^2)$, $v_y \textapprox \mathcal{N}(0,0.06^2)$, $v_{\dot{x}} \textapprox \mathcal{N}(0.0,0.04^2)$, $v_{\dot{y}} \textapprox \mathcal{N}(0.0,0.04^2)$ and $w_t=0$. Fig.~\ref{fig:robot_corr} presents
the static obstacles, goal, and the reference trajectory. To increase the fidelity of the simulation, we add intermediate points as discussed in Section~\ref{sec:coll_avoid}. In the first scenario, the horizon $T=5$sec with $\Delta t=1$sec. The STL specification:
\begin{equation}
\label{eqn:prob_robot_corr_spec}
\varphi_1 := \phi_{\mathit{0}} \wedge \Box_{[0,5]} \neg \phi_{\mathit{obs1}} \wedge \Diamond_{[5,5]} \phi_{\mathit{goal}} 
\end{equation}
Where $\phi_z=\operatorname{True}$ if the intersection of all predicates of $z$ over the state $x$ is greater than zero.
In this case, $\phi_{obs1}$ is non-convex thus we use Delaunay triangulation \cite{lee1980two} to decompose it into two convex polytopes $\phi_{\mathit{obs1:1}},\phi_{\mathit{obs1:2}}$. We have not considered any other restrictions on the state except on the pose. %

\subsubsection{Setup}
We compute $p(fail)=p(\neg\varphi_1)$. We construct a disjunction between the trajectory-space $\mathcal{H}$-polytopes of failing trajectories of type $\mathcal{L}(\varphi_a,0)$ and $\mathcal{L}(\varphi_b,0)$ as discussed in Section~\ref{sec:coll_avoid}. %

\subsubsection{Results}
\begin{figure}
     \centering
     \begin{subfigure}[b]{0.495\columnwidth}
         \centering
         \includegraphics[width=\columnwidth, height=3.2cm]{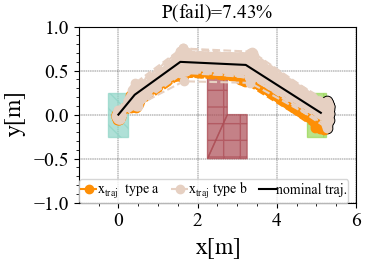}
         \caption{Scenario $1$: $\Delta t=1$sec.\\ $p(fail)=7.43\%\pm0.8\%$.}
         \label{fig:robot_corr}
     \end{subfigure}
     \hfill
     \begin{subfigure}[b]{0.495\columnwidth}
         \centering
         \includegraphics[width=\textwidth, height=3.2cm]{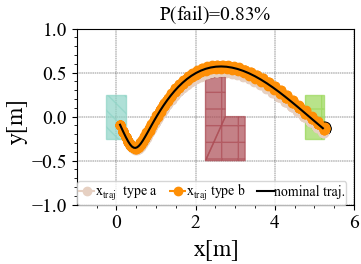}
         \caption{Scenario $2$: $\Delta t=0.1$sec.\\ $p(fail)=0.83\%\pm0.26\%.$}
         \label{fig:robot_corr_large}
     \end{subfigure}
     \caption{X-Y projection of a robot maneuvering in a field with obstacles (red) and goal (green). In orange, the failing trajectories of type ($a$). In gray, the failing trajectories of type ($b$). In black, the reference trajectory.}
     \label{fig:robot_nav}
     \Description{In figure a, a 2-d projection of the robot's workspace. Trajectories examples of type a and type b start from the initial set on the left, all the way to the goal set on the right. There is an obstacle in the middle. The time step is 1 sec, horizon is 5 sec, meaning 5 step horizon meaning 20-dimension  trajectories. The probability to fail is 10.5\%. Figure b is the same, only for a time step of 0.1sec, for a 200-dimension of a trajectory. The probability to fail is 3.9\%.}
\end{figure}

Fig.\ref{fig:robot_corr} shows a sample of the trajectories of both failure modes. Computing the probability with our proposed algorithm yields $p(fail)=7.43\%\pm 0.8\%$ and took $25$sec. Verifying the same system with Monte-Carlo simulations yields $p_{MC}(fail)=7.66\% \pm 0.5\%$ and took $13.0$sec. To estimate the probability of failing with Monte-Carlo, we use $n_{MC}=2400$ simulations, where $\sigma^2 = \hat{p}(1-\hat{p})/n_{MC}$. The experiments were done on an Intel(R) Core(TM) i7-6700 and $n_k=n=256$ samples in each nesting of the HDR algorithm (to get a comparable standard deviation). Monte-Carlo yields faster results because the probability to fail is relatively high.

\begin{figure}
     \centering
        \includegraphics[width=\columnwidth, height=5cm,keepaspectratio]{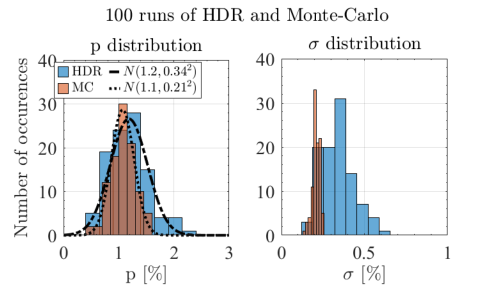}
     \caption{Running our approach (HDR) and MC (mean of $2400$ runs) a 100 times each, and the comparison of the mean $\bar{p}(fail)$ estimation and its standard deviation.}
     \label{fig:mc_of_hdr_and_mc}
     \Description{Figure on the left is a histogram of a 100 runs of Monte-Carlo and 100 runs of our method. It shows that on average, both techniques yield the same results p=2.5 in the robot navigation problem (for the 200-dimension trajectories). On the right, a histogram of the variances of each run and each method. The difference in the mean is 0.1\%. The width of our method in this case is slightly bigger.}
\end{figure}

Using a horizon of $5$ steps yields a simulation of $20$ random variables which can be considered a relatively small parameter space. Fig.\ref{fig:robot_corr_large} depicts the second scenario with similar settings (initial conditions were changed to induce failures because the LQR controller with the new time step performs differently) for a horizon of $T=5$sec and $\Delta t=0.1$sec. This time, the problem dimension is $200$. With our algorithm, $p(fail)=0.83\%\pm0.26\%$ and took $110$sec to complete with $n_k=64$. Monte-Carlo simulation takes $327$sec and yields $p_{MC}(fail)=1.2\%\pm0.2\%$ in $2400$ simulations. Fig.\ref{fig:mc_of_hdr_and_mc} shows the distribution of running our algorithm and MC $100$ times with different seeds and the results match for $\bar{p}(\neg\varphi_1)$, $\sigma$ depends on $n_k$.

\subsection{Car passing an intersection - reach-avoid}
\label{sec:demo_car_intersection}

\subsubsection{Setup}

\begin{figure}[h]
     \centering
     \begin{subfigure}[b]{0.4\columnwidth}
         \centering
         \includegraphics[width=\textwidth]{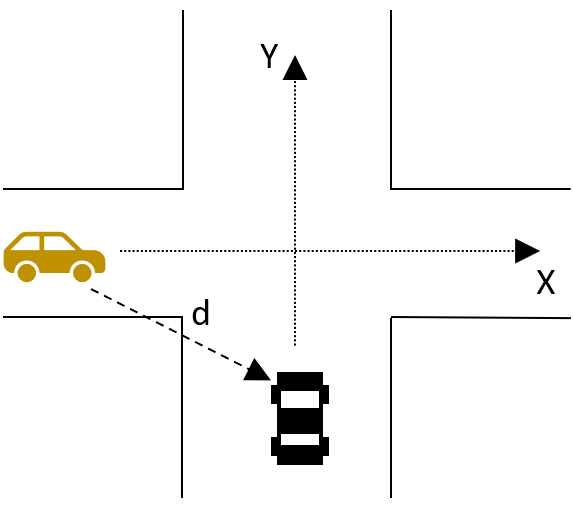}
         \caption{}
         \label{fig:ex_car_view}
     \end{subfigure}
     \hfill
     \begin{subfigure}[b]{0.48\columnwidth}
         \centering
         \includegraphics[width=\textwidth]{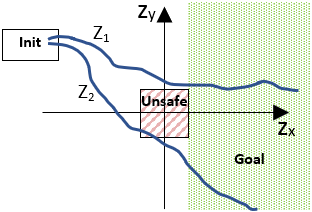}
         \caption{}
         \label{fig:ex_car_trajs}
     \end{subfigure} \\
     \begin{subfigure}[b]{0.48\columnwidth}
         \centering
         \includegraphics[width=\textwidth]{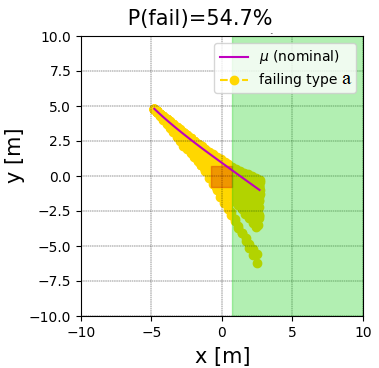}
         \caption{}
         \label{fig:intesection_trajectories}
     \end{subfigure}
     \setlength{\belowcaptionskip}{-7pt}
     \caption{(a) Controlled car (yellow, left) entering an intersection with an uncontrolled car (black, bottom). (b)  A satisfying trajectory ($z_1$), and a violating trajectory ($z_2$) in the Z coordinate frame. (c) A single noise sequence example of failing trajectories that intersect with the unsafe set.}
     \label{fig:ex_car_inters}
     \Description{In figure (a), a schematic of an road intersection, a controlled orange car tries to cross from left to right, on the bottom a black uncontrolled car trying to reach the top. In figure (b), the 2-D representation of the workspace with the problem's state vector, the initial set, the unsafe set about the origin and the goal set on the right of the intersection, and some schematic representations of possible trajectories. In figure c, the results of our technique with sample trajectories that fail to satisfy the specification overlaid on the unsafe zone and goal zone.}
\end{figure}

We consider a controlled car (Ego-vehicle, $E$) driving along the $x$-axis and an uncontrolled car (Other vehicle, $O$) driving along the $y$-axis, as shown in Fig.~\ref{fig:ex_car_view}. Each car's dynamic equations follow the holonomic robot in \eqref{eqn:holonom_dyn_approach}.
Since $O$ is uncontrolled we model its dynamics with a process noise $w_{\dot{y}}^O \textapprox \mathcal{N}(0,0.2^2)$.
$E$ is measuring the distance $d$ to $O$ using a Lidar and has errors \cite{thrun2002probabilistic}. The error modes are a Gaussian about the true value, and a maximum range error $v_d^E \textapprox \pi_{1}\mathcal{N}(0,0.04^2)+\pi_{2}\mathcal{N}(5,0.6^2)$. The transitions between the Gaussians are expressed with a Markov chain $p(\pi_1(t)\given \pi_1(t-1))=0.98$, $p(\pi_2(t)\given \pi_1(t-1))=0.02$, $p(\pi_1(t)\given \pi_2(t-1))=0.6$, $p(\pi_2(t)\given \pi_2(t-1))=0.4$ indicating the probability of having a bad measurement after a previous bad measurement is higher (occlusion, multipath). $E$ needs to cross the intersection safely and uses the control law: $u^E_t=u_0 - Kd=u_0 - K (\sqrt{(x^O_t-x^E_t)^2-(y^O_t-y^E_t)^2}+v_d)$. The time horizon is $T=3$sec and $\Delta t=0.1$sec. The cars' lengths are $L=1.0$m and widths $W=0.5$m. $K=-0.1$ and $u_0=0.075$ such that when the distance between the cars $d \leq 0.5(L+W)$, the control yields $u_k=0$ and $E$ stops until $O$ crosses the intersection. 

We derive a new state variable $z=[x^E_k-x^O_k,y^E_k-y^O_k,\dot{x}^E_k-\dot{x}^O_k,\dot{y}^E_k-\dot{y}^O_k]'$ with the initial conditions $z_0=[-5,5,2,-2]'$, as shown in Fig.~\ref{fig:ex_car_view}. In this new state variable, it is easy to show that the unsafe set (the ``obstacle'') is a square centered at the origin of $z_x, z_y$ where the lengths of all the sides are $L+W$. The goal is for $E$ to cross to the other side, $z_x \geq 0.5(L+W)$. The polytope sets are shown in Fig.~\ref{fig:ex_car_trajs}. The STL specification:
\begin{equation}
\label{eqn:prob_intersection_spec}
\varphi_{int} := \Box_{[0,3]} \neg \phi_{\mathit{unsafe}} \wedge \Diamond_{[2.9,3]} \phi_{\mathit{goal}} 
\end{equation}
Here, the measurement equation is non-linear. We find the trajectory's Gaussian distribution by linearizing the distance measurement in \eqref{eqn:prob_intersection_linearization} evaluated at the expected value of the state, \eqref{eqn:prob_intersection_expectedz}.
\begin{equation}
\label{eqn:prob_intersection_linearization}
C_t=\frac{\partial d}{\partial z}\given_{z=E[z_t]} = \frac{1}{d}[z_x, z_y, 0, 0] 
\end{equation}
\begin{flalign}
\label{eqn:prob_intersection_expectedz}
\EX[z_{t+1}] = & (A-BKC_t)\EX[z_t] + Bu_0+\EX[w_t^O]   -BK\EX[v_t^E]  
\end{flalign}

\subsubsection{Results}
Fig.~\ref{fig:intesection_trajectories} shows the failing trajectory samples of one iteration of sampled noise mixing factors (Sec.~\ref{sec:guassian_mix}). Total probability to fail $p(fail)=54.68\%$ with $95\%$ confidence level $[52.73\%, 56.63\%]$. We compare with Simple Random Sampling (SRS) with $n_{MC}=2500$ for the Monte-Carlo simulations of the full non-linear system. The probability estimate $p_{MC}(fail)=54.08\% \pm 1.0\%$ took $T=32$sec to run, while using our method took $T=130$sec (again, the times are due to the high probability of failure).

\subsection{Data-based simulation - reach-avoid}
\begin{figure}
     \centering
     \begin{subfigure}[b]{0.48\columnwidth}
         \centering
         \includegraphics[width=\textwidth, height=3cm,keepaspectratio]{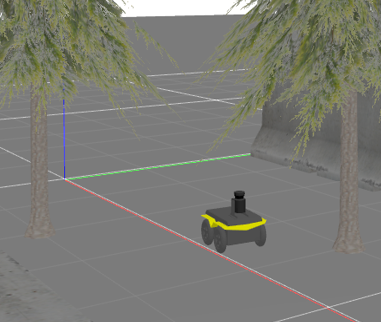}
         \label{fig:robot_gazebo}
     \end{subfigure}
     \hfill
     \begin{subfigure}[b]{0.48\columnwidth}
         \centering
         \includegraphics[width=\textwidth,, height=3cm,keepaspectratio]{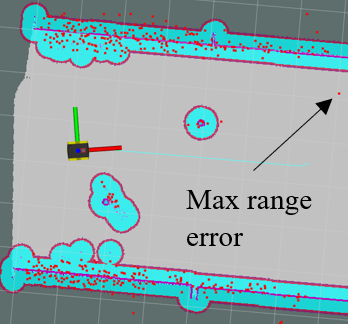}
         \label{fig:robot_rviz}
     \end{subfigure}
     \caption{Jackal navigating in the environment (left). Lidar measurements and robot's current mapping (right).}
     \label{fig:robot_realistic_scenario}
     \Description{In figure (a) a picture of a Jackal robot in simulation with trees. In figure (b), a view showing the robot, its Lidar sensing locations and its internal knowledge of the map with a distinct indication of a wrongly sensed Lidar ray (maximum range error).}
\end{figure}

\noindent In this example we show how this technique can be used in a scenario where the noises or system dynamics are not known. For this demonstration we run the Jackal \cite{jackal} robot in the Gazebo simulator \cite{gazebo:2020} with the Robot Operating System (ROS) \cite{ROS}. We use the built-in controllers and estimation algorithms provided with and for the robot, and send it a goal command. With probability of $5\%$, a maximum range noise \cite{thrun2002probabilistic} is injected to any of the Lidar's ray measurements. Fig.~\ref{fig:robot_realistic_scenario} shows the environment the robot is navigating through. Our purpose is to verify that the system can reach the goal safely. However, running a single run takes approximately $15$sec, thus making a Monte-Carlo simulation intractable when the failure rate is low. 

In our approach, we first run $n$ simulations and fit a multivariate Gaussian (e.g. \textit{robustcov} in Matlab) to the set of (ground truth) trajectories. $n$ must be at least twice the number of variables (states x time steps). We now have $x_{traj}\sim \mathcal{N}(\mu,\Sigma)$ and we directly compute the probability to collide with a tree, miss the goal or violate any other temporal constraint. 

\begin{figure}
     \centering
        \includegraphics[width=\columnwidth, height=4cm,keepaspectratio]{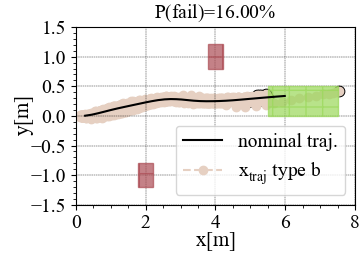}
     \caption{Data-based verification - $x^\varphi$ comes from simulations. $16.0\%$ of the trajectories are not at the goal at $13.2$sec.}
     \label{fig:gazebo_failing_trajs}
     \Description{In the figure, a 2-d view of the workspace of the Gazebo example with the results. Only type b failures are available, and they show how the goal set is not reached in the specification time-frame.}
\end{figure}

In Fig.~\ref{fig:gazebo_failing_trajs} we show the verification results for the system with %
\begin{equation*}
\varphi_{Gazebo}:=\Box_{[0,13.2]}\left( \neg \operatorname{Tree}_1 \wedge \neg \operatorname{Tree}_2 \right) \wedge \Box_{[13.2,13.2]}\operatorname{Goal}
\end{equation*}
 where $\operatorname{Goal}$ is the region defined by the box $5.5 \leq x \leq 7.5, ~0 \leq y \leq 0.5$. The time step in this scenario is $\Delta t=0.4$sec and a horizon of $13.2$sec. We see that $16.0\%$ of the trajectories fail to reach the goal on time (or overshoot it). We stopped the computation of the probability of hitting a tree (``$\varphi_a$'') at $k=24$ nestings, which means that a crash is less likely than about $6\cdot10^{-6}\%$.

\subsection{Robot navigation - Full STL}
\label{sec:full_stl_demo}
\begin{figure}
    \centering
     \begin{subfigure}[b]{0.48\columnwidth}
         \centering
         \includegraphics[width=\columnwidth]{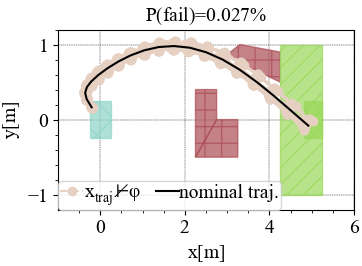}
         \caption{}
         \label{fig:full_stl_ex_hdr}
     \end{subfigure}
     \hfill
     \begin{subfigure}[b]{0.48\columnwidth}
         \centering
         \includegraphics[width=\columnwidth,keepaspectratio]{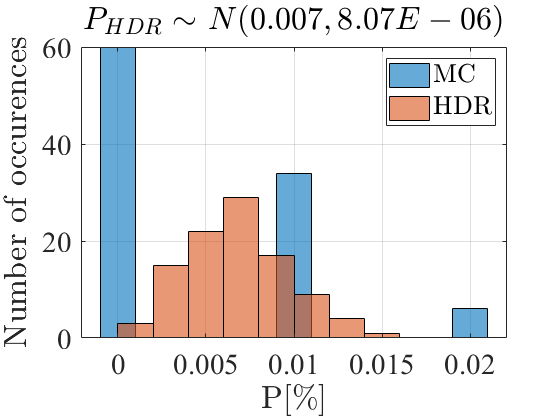}
         \caption{}
         \label{fig:full_stl_ex_hdrvsmc}
     \end{subfigure}
     \caption{(a) Failing trajectories of the STL formula $\varphi_{STL}$. (b) Statistics for a 100 trials for HDR, and for 100 MC, each with $10^4$ simulations.}
     \label{fig:full_stl_ex_hdr_all}
     \Description{In figure (a), the results and sample trajectories that fail the specification of a full STL specification. There are several complex obstacles and two goal regions and an initial set involved in the specification. In figure (b), a histogram of a 100 runs with Monte-carlo (each with 10000 simulations) and a 100 runs with our method. We see that most runs of Monte-Carlo yield no failing examples, while our method shows it every single time. Also, our method takes half the time to run without optimizations.}
     
\end{figure}

\noindent In this example we consider the robot in Example \ref{example_holonomic_robot} and a complex STL specification: 
\begin{flalign}
\varphi_{STL}:= & \Box_{[0,T]} \neg \left( \phi_{\operatorname{Obs}_1} \vee \phi_{\operatorname{Obs}_2} \right) \wedge \\\nonumber & \Box_{[0,T]} (\phi_{\operatorname{Goal}_1} \implies \Diamond_{[0,0.25]} \phi_{\operatorname{Goal}_2} )
\end{flalign}
where $(a\implies b) = (\neg a \vee b) $. %
Since our technique computes the probability of \textit{satisfying} the STL formula, to find the probability of failure, we use $\neg\varphi_{STL}$ in our computations. Following a single run of our method, we are able to find violating trajectories (Fig.~\ref{fig:full_stl_ex_hdr}) even though  $p(fail)=0.027\%$. To find just one event with this probability we would need to run approx. $4000$ simulations with MC. We ran $100$ trials with our technique, and $100$ trials with MC with $10^4$ simulations each. The results are shown in Fig.~\ref{fig:full_stl_ex_hdrvsmc}. $60\%$ of the MC runs end with no failing examples, and about a third end with one failing example. The mean time to run MC is $626\pm10$sec and our method is $333\pm35$sec. The minimal probability computed by our method is $p(fail)=0.001\%$.

\begin{figure}
     \centering
        \includegraphics[width=\columnwidth,height=4cm,keepaspectratio]{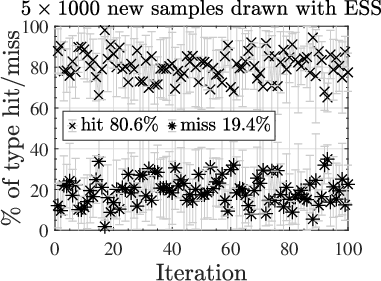}
     \caption{The assignation between trajectories that fail due to obstacle collision and due to missing the goal in time. We collect five $\times1000$ new trajectories with ESS after each round of $p(\varphi_{STL})$ computation. The markers are the means, and the error bars are $1$ standard deviation.}
     \label{fig:ess_samples_analysis}
     \Description{A scatter plot that shows the mean number of failure types (a failure that has to do with the obstacles, or a failure that has to do with the goals) that are sampled within the union of all domains. We run the approach for a hundred times, and then sample 5 times a 1000 new trajectories from the domain of falsifying trajectories. The plot also shows the variance of the trajectory types sampled with error bars.}
\end{figure}

Due to the use of the STL score for full STL, one cannot identify the specific cause of the failure. %
Furthermore, the sampled trajectories that fail the specification do not necessarily represent the proportions of the different failure causes.   
This is due to two reasons - first, we cannot guarantee how many trajectories are present in the final nesting, as explained in Section~\ref{sec:Preliminaries}. Second, because this is a MCMC approach, the samples might be biased towards a certain region given an initial sample within that region. However, we show that if we sample new trajectories, we will get the correct proportions on average given enough samples. For example, in the previous scenario, the ratio between the probability mass for hitting the obstacle at $t_{14}$ and not making the second goal on time, is approximately 4:1. We ran our approach $100$ times. After each iteration finished, we sampled five sets of $1000$ samples that violate $\varphi_{STL}$ and computed how many of those hit the obstacle and how many violated the goal requirement. Results are shown in Fig.~\ref{fig:ess_samples_analysis}; although at specific instances we can even get more trajectories of goal violations than obstacle violations, we see that on average, we sample the correct proportions. This means that with our method, we are able to ``jump'' from an active domain to another active domain even if it is clearly distinct (different predicates and different time bounds). Of course, regions may be overshadowed by regions with considerably higher probability mass and if one wants to check those too, then they might need to decompose the specification to capture only those.

\subsection{Adversarial Scenarios - Full STL}
\label{sec:demos_adv_scn}
\begin{figure}
     \centering
        \includegraphics[width=\columnwidth,height=4cm,keepaspectratio]{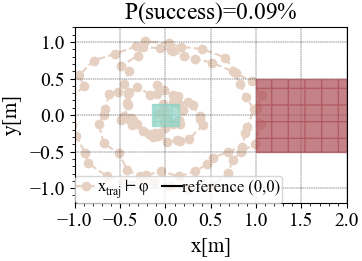}
     \caption{Discovering adversarial noise sequences that lead to unsafe behaviors.}
     \label{fig:guided_synth_ex}
     \Description{A 2-D view of the workspace of an example from the paper reference 30. A system with a regulator aimed at keeping it in the origin is swept away towards the unsafe zone (a rectangle [1,2]x[-0.5x-0.5]) with a series of noises that we can detect with our method. The method can find different trajectories coming from different predicates and different time steps (different linear trajectory-level constraints) such that eventually the system hits the unsafe set.}
\end{figure}

\noindent In \cite{yang2021synthesis} the authors developed a synthesis guided approach to find adversarial examples that falsify a dynamical system with respect to reach-avoid type specifications. An example from that paper (\textit{Example 2}) finds a series of measurement noises that causes the system \eqref{eqn:guided_synth_syseqn} and its regulator to enter the unsafe zone. 
\begin{flalign}
\label{eqn:guided_synth_syseqn}
\xi_{t+1}&=
\begin{bmatrix}
    0.9745 & 0.2132 \\
    0.002547 & 1.151
\end{bmatrix} \xi_t+ 
\begin{bmatrix}
    0.01959 \\
    0.1961 
\end{bmatrix} u_t + 
\begin{bmatrix}
    0.01959 \\
    -0.04509 
\end{bmatrix} 
w_t \\\nonumber
u_t&= -\begin{bmatrix} 1 & 1 \end{bmatrix} \eta_t ~;~
\eta_{t}= \xi_t+v_t
\end{flalign}
In \cite{yang2021synthesis}, the noises $v_t=[-0.1,0.1]^2$ and $w_t=[-0.2,0.2]$ are uniform and bounded. Here we approximate them with an appropriate Gaussian. The unsafe set is defined $\operatorname{Unsafe}(\xi)=[1,2]\times[-0.5,0.5]$ and the system starts in $\operatorname{Init}(\xi)=[-0.15\times 0.15]^2$. To find adversarial trajectories, we consider the STL formula:
\begin{flalign}
\label{eqn:guided_synth_spec}
\varphi_{adv} := \operatorname{Init}(\xi) \wedge \Diamond_{[0,115]}\operatorname{Unsafe}(\xi)
\end{flalign}
In addition to finding the probability, our approach can find adversarial examples, as done in \cite{yang2021synthesis}. In Fig.~\ref{fig:guided_synth_ex} we show a trajectory, sampled from the set of satisfying trajectories, that eventually enters the unsafe zone. %
In this example, the probability that the system may enter the unsafe zone is $0.09\%$ where the different trajectories may enter the unsafe zone at different times; our apporach can provide several such examples.  %

\section{Discussions and Conclusions}
\label{sec:discuss}

\noindent In this paper we introduced a method to accurately compute the probability that a linearizable system 
will satisfy an STL specification. The framework is general and can accommodate various sensor, estimator and perception errors. We provide two methods for calculating the probability - for full STL and for reach-avoid specifications. %

Our method, while including computation overhead, is scalable to high dimensions (longer horizons or models with more states) and its computational complexity does not depend on the combinatorially many solutions of the specification. The sampling is efficient, especially in low probability events where a naive Monte-Carlo approach may not be tractable. The latter may suffer from dimension explosion, leading to the need for a large number of simulations to adequately sample the posterior. Our method is sampling from the posterior in a rejection-free and parameter-free (no hyper parameters needed for the slice sampler) manner.

Our method lends itself to parallel implementation, thereby reducing the computation time. %
Every nesting from the ESS and HDR can be run in parallel. 
By increasing the computation speed, %
our method can potentially be used as a step in motion planning, where, for example, we can check the output of a rapidly exploring random tree (RRT) generated path to check feasibility given noises or complex specifications.

In future work we will use this framework to synthesize controllers that can minimize the probability of failure. Another direction is to use the sampled failed trajectories to gain insight and requirements on the perception system that would be the most beneficial in reducing failure. %

\begin{acks}
 This work is supported by \grantsponsor{PERI_MURI}{ONR PERISCOPE MURI} a award \grantnum{PERI_MURI}{N00014-17-1-2699}.
\end{acks}

\bibliographystyle{ACM-Reference-Format}
\bibliography{refs}
\end{document}